\documentclass[twocolumn,aps,showpacs,prb,tightenlines,amsmath,amssymb,superscriptaddress]{revtex4}

\usepackage{graphicx}

\usepackage{dcolumn}

\usepackage{bm}

\usepackage{colordvi}

\begin{document}

\title{Detection of large magneto-anisotropy of electron spin dephasing
in a high-mobility two-dimensional electron system in a $[001]$
GaAs/AlGaAs quantum well}

\author{D.\ Stich}
\affiliation{Institut f\"ur Experimentelle und Angewandte Physik,
Universit\"at Regensburg, D-93040 Regensburg, Germany}
\author{J.\ H.\ Jiang}
\affiliation{Hefei National Laboratory for Physical Sciences at
Microscale and Department of Physics, University of Science and
Technology of China, Hefei, Anhui, 230026, China}
\author{T.\ Korn}
\affiliation{Institut f\"ur Experimentelle und Angewandte Physik,
Universit\"at Regensburg, D-93040 Regensburg, Germany}
\author{R.\ Schulz}
\affiliation{Institut f\"ur Experimentelle und Angewandte Physik,
Universit\"at Regensburg, D-93040 Regensburg, Germany}
\author{D.\ Schuh}
\affiliation{Institut f\"ur Experimentelle und Angewandte Physik,
Universit\"at Regensburg, D-93040 Regensburg, Germany}
\author{W.\ Wegscheider}
\affiliation{Institut f\"ur Experimentelle und Angewandte Physik,
Universit\"at Regensburg, D-93040 Regensburg, Germany}
\author{M.\ W.\ Wu}
\email{mwwu@ustc.edu.cn.} \affiliation{Hefei National Laboratory
for Physical Sciences at Microscale and Department of Physics,
University of Science and Technology of China, Hefei, Anhui,
230026, China}
\author{C.\ Sch\"uller}
\email{christian.schueller@physik.uni-regensburg.de.}
\affiliation{Institut f\"ur Experimentelle und Angewandte Physik,
Universit\"at Regensburg, D-93040 Regensburg, Germany}

\date{\today}

\begin{abstract}
In time-resolved Faraday rotation experiments we have detected an
inplane anisotropy of the electron spin-dephasing time (SDT) in an
$n$--modulation-doped GaAs/Al$_{0.3}$Ga$_{0.7}$As single quantum
well. The SDT was measured with magnetic fields of $B\le 1$ T,
applied in the $[110]$ and $[1\bar{1}0]$ inplane crystal directions
of the GaAs quantum well. For fields along $[1\bar{1}0]$, we have
found an up to a factor of about 2 larger SDT than in the
perpendicular direction.  The observed SDTs also show strong
anisotropic magnetic field dependence.
 Fully microscopic calculations, by
numerically solving the kinetic spin Bloch equations considering the
D'yakonov-Perel' and the Bir-Aronov-Pikus mechanisms, reproduce the
experimental findings quantitatively. This quantitative analysis of
the data allowed us to determine the relative strengths of Rashba
and Dresselhaus terms in our sample. Moreover, we could
predict the SDT for spins aligned in the $[110]$ {\em inplane}
direction to be on the order of several nanoseconds, which is up to
two orders of magnitude larger than that in the perpendicular {\em
inplane} direction.
\end{abstract}


\maketitle

The exploration of electron spin relaxation and dephasing is at the
heart of the research of semiconductor
spintronics.\cite{Awschalom1,Fabian} For potential applications in
quantum computation or spin transistor devices,\cite{DattaDas} the
investigation and knowledge about the relevant spin dephasing
channels is an important prerequisite. The most powerful techniques
for the study of spin dynamics in semiconductors are all-optical
techniques, like time-resolved photoluminescence or time-resolved
Kerr or Faraday rotation (TRFR), where the latter are so called
pump-probe techniques. In fact, TRFR experiments have helped to
detect extremely long spin relaxation times in bulk GaAs structures,
which were $n$-type doped, close to the metal-to-insulator
transition.\cite{Kikkawa1} So far, however, comparatively little
attention has been paid to the experimental investigation of spin
dynamics in quantum wells containing free electrons ({\em e.g.}, in
Refs.\ \onlinecite{Kikkawa2,Richard1,Richard2}). In the majority of
experiments, spin-aligned charge carriers are created via the
absorption of circularly-polarized light. Hence, the direction of
optically-oriented spins is parallel or antiparallel to the
direction of light propagation. Therefore, in typical time-resolved
experiments on quantum-well samples with normal incidence, the spins
of the photoexcited carriers point perpendicular to the quantum-well
plane, {\em i.e.}, in growth direction of the layered sample. In a
pioneering experiment, Ohno {\em et al.}\cite{Ohno,Dohrmann}
confirmed experimentally in quantum-well samples, which were grown
in the $[110]$ crystal direction, a very long spin relaxation time
on the order of nanoseconds for spins, aligned parallel to the
growth direction. This was predicted theoretically to be due to the
absence of the D'yakonov-Perel' spin relaxation mechanism\cite{dp}
in this particular crystal direction. These discoveries lead to a
growing interest in $[110]$ structures, which promised to be
well-suited candidates for spintronics' devices. There are, however,
essentially two important reasons, which might suggest to use
standard $[001]$-grown heterostructures for spin-injection devices:
(i) In typical real transport devices for spin injection and
manipulation, as, {\em e.g.}, motivated by the Datta-Das spin
transistor,\cite{DattaDas} layered ferromagnetic electrodes are
employed as spin injectors, which exhibit inplane magnetization
directions, {\em i.e.}, the injected spins are typically {\em
inplane} with respect to the semiconductor structure. (ii) Due to
the lower growth temperature for the growth on $[110]$ substrates,
the mobility in modulation-doped heterostructures, grown in that
crystal direction, is generally lower than that in standard $[001]$
structures.

Quite some time ago it was demonstrated theoretically that also in
$[001]$ semiconductor heterostructures of Zincblende type, spin
relaxation could be greatly suppressed for spins pointing into the
$[110]$ {\em inplane} direction.\cite{Averkiev} A giant spin
relaxation anisotropy was predicted for inplane spin orientation,
where spin relaxation should be strong in $[1\bar{1}0]$ direction
and weak in the perpendicular $[110]$ direction. The strength of the
anisotropy roughly depends on the ratio between the Rashba
\cite{Rashba} and Dresselhaus \cite{dress} spin-orbit coupling
terms: The anisotropy should be maximal if both terms have equal
strengths.\cite{cheng} For this case, Schliemann {\em et al.}
proposed a spin transistor device which uses diffusive
transport.\cite{Schliemann03} Surprisingly, a strongly reduced spin
relaxation in the $[110]$ inplane direction in two-dimensional
electron systems (2DES) in standard $[001]$-grown heterostructures,
so far has not been detected in a time-resolved experiment.
Recently, Averkiev {\em et al.} performed polarization-resolved cw
photoluminescence measurements on a nominally undoped $[001]$-grown
quantum well at liquid Nitrogen temperature,\cite{Averkiev06}
determining the depolarization of the photoluminescence in an
applied inplane field (Hanle effect). From the experimental data the
authors extracted inplane spin relaxation times, which differed by
about a factor of 2 for the two perpendicular inplane directions,
which showed evidence for the inplane anisotropy of the SDT. More
recently,
  Liu {\em et al.} performed Kerr rotation measurement on a
[001]-grown two dimensional electron gas at 150\ K and extracted
 the inplane SDTs along different directions with a factor of 1.3.
In this work we report TRFR experiments on a high-mobility 2DES with
inplane magnetic fields. We could detect a magneto-anisotropy of the
SDT, measured via the out-of-plane component of the initial spin
polarization. Calculations employing the fully microscopic kinetic
spin Bloch equations,\cite{wu,wu1,wu2,wu3,cheng1} taking into
account {\em all} relevant spin relaxation mechanisms (including
both the D'yakonov-Perel' and
 the Bir-Aronov-Pikus mechanisms\cite{BAP,zhou}), reproduce the experimental data
quantitatively without any parameter for the SDT. This allows us
to estimate the maximum inplane SDT in our standard high-mobility
quantum-well structure to be on the order of several nanoseconds,
which is about 2 orders of magnitude larger than in the
perpendicular inplane direction.

\begin{figure}[t]
\begin{center}\includegraphics[width=8.5cm]{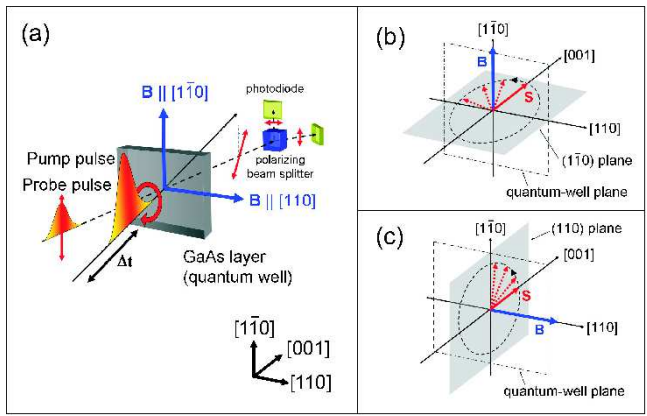}\end{center}
\caption{(color online) (a) Schematical picture of the
time-resolved Faraday rotation experiment. Inplane magnetic fields
are applied either in the $[110]$ or the $[1\bar{1}0]$ directions.
(b) Sketch of the precession of optically oriented spins about a
$[110]$ inplane magnetic field. (c) Same as (b) but for
$[1\bar{1}0]$ magnetic field. }\label{Fig1}
\end{figure}

Our sample is a 20\ nm-wide, one-sided modulation-doped
GaAs-Al$_{0.3}$Ga$_{0.7}$As single quantum well. The electron
density and mobility at $T=4.2$\ K are $n_e=2.1\times 10^{11}$\
cm$^{-2}$ and $\mu_e=1.6\times 10^6$\ cm$^2$/Vs, respectively. For
measurements in transmission geometry, the sample was glued onto a
glass substrate with an optical adhesive, and the substrate and
buffer layers were removed by selective etching. The sample was
mounted in the $^3$He insert of a superconducting split-coil magnet
cryostat. All measurements were taken at a temperature of $T=4.5$ K.
For the TRFR measurements, two laser beams from a mode-locked
Ti:Sapphire laser, which is operated at 80\ MHz repetition rate,
were used. The laser pulses had a temporal length of about 600\ fs
each, resulting in a spectral width of about 3-4\ meV, which allowed
for a resonant excitation. The laser wavelength was tuned to excite
electrons from the valence band to states slightly above the Fermi
energy of the host electrons in the conduction band. Both laser
beams were focused to a spot of approximately 60\ $\mu$m diameter on
the sample surface. The pump pulses were circularly polarized in
order to create spin-oriented electrons in the conduction band, with
spins aligned perpendicular to the quantum well plane, {\em i.e.},
in $[001]$ direction. The pump power was set to excite an initial
spin polarization of the 2DES of about 6\%. Such a high spin
polarization ensures a long spin dephasing time to subtract the spin
dephasing time from the full spin precession at low magnetic
field.\cite{wu1,Stich} The rotation of the linear polarization of
the time-delayed probe pulse, due to the Faraday effect, was
measured by an optical bridge (see Fig.\ \ref{Fig1}a). Due to its
near-normal incidence on the sample, the polarization rotation of
the probe pulse is caused by the out-of-plane component of the spin
polarization of the 2DES.

Figure\ \ref{Fig1}a is a schematic of the experiment, showing the
orientation of the sample, relative to the laser beams and the
magnetic fields. Figures\ \ref{Fig1}b and \ref{Fig1}c illustrate the
precessional motion of the spin polarization in the two cases
investigated in this paper: For a magnetic field applied parallel to
the $[1\bar{1}0]$/$[110]$ direction (Fig.\ \ref{Fig1}b/\ref{Fig1}c),
the spins, which are created with an out-of-plane orientation
parallel to the $[001]$ direction, are forced to precess within the
$(1\bar{1}0)$/$(110)$ \emph{plane}. Besides the magnetic field, the
spins also precess  around the momentum ${\bf k}$-dependent
effective magnetic field due to the Dresselhaus and Rashba
spin-orbit couplings which provide an inhomogeneous broadening
leading to the spin dephasing.\cite{wu,wu1,wu2,wu3,cheng1,cheng}
Spins, precessing within the $(1\bar{1}0)$/$(110)$ plane, feel the
inhomogeneous broadening along the $[1\bar{1}0]$/$[110]$ direction,
which reads $(\alpha-\beta)k_x$/$-(\alpha+\beta)k_y$, if the
$y$/$x$-axis is along $[1\bar{1}0]$/$[110]$.\cite{cheng} Here,
$\alpha$/$\beta$ represents the Rashba/Dresselhaus coefficient.
Strong anisotropy is predicted when $\alpha$ is comparable to
$\beta$. Thus, during their precessional motion, the spins are
expected to probe the relaxation times in the different inplane
directions. In this work we demonstrate that the effect on their
overall, averaged SDT can be observed by measuring the out-of-plane
component of the spin polarization.

\begin{figure}[t]
\begin{center}\includegraphics[width=7cm]{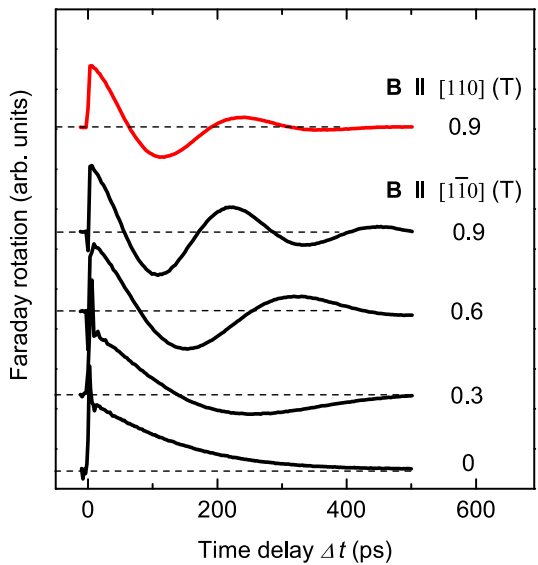}\end{center}
\caption{(color online) Experimental TRFR traces for different
directions of the inplane magnetic field.}\label{Fig2}
\end{figure}

\begin{figure}[t]
\begin{center}\includegraphics[width=8.5cm]{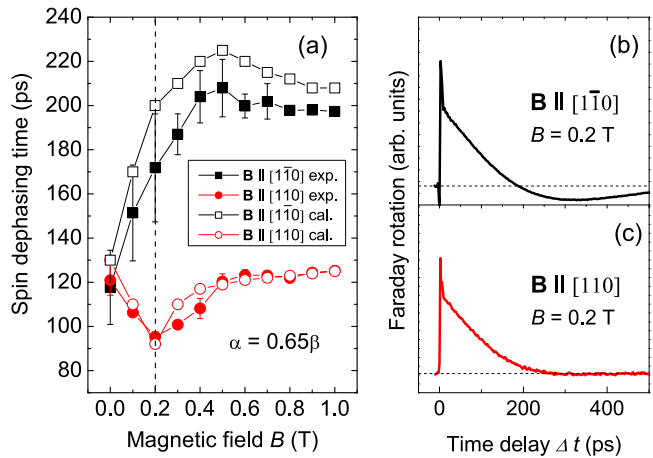}\end{center}
\caption{(color online) (a) Comparison of experimental (solid
symbols) and theoretically calculated (open symbols) spin
dephasing times for different inplane magnetic-field directions.
$\alpha$ and $\beta$ are the Rashba and Dresselhaus spin-orbit
coefficients, respectively. (b),(c) Comparison of experimental
TRFR traces at $B=0.2$\ T for the two different inplane directions
of the magnetic field.}\label{Fig3}
\end{figure}

\begin{figure}[t]
\begin{center}\includegraphics[width=7.5cm]{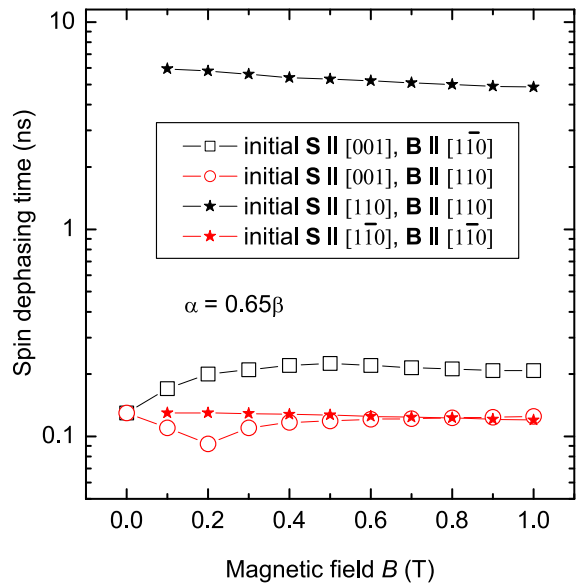}\end{center}
\caption{(color online) Comparison of spin dephasing times,
calculated with identical parameters for initial spin
polarizations parallel to the $[001]$ (open symbols) and the
$[110]$ and $[1\bar{1}0]$ directions (solid stars) in dependence
on magnetic fields parallel to the $[110]$ and $[1\bar{1}0]$
inplane directions. }\label{Fig4}
\end{figure}

Figure\ \ref{Fig2} shows TRFR traces taken for various
magnetic-field amplitudes applied parallel to the $[1\bar{1}0]$
direction (black traces), which show an exponentially damped
oscillation of the out-of-plane component of the spin polarization.
The temporal decay of the signal is determined by the SDT. A TRFR
trace taken for a magnetic field of 0.9\ T parallel to $[110]$ (red
trace) is shown directly above for comparison. By comparing it to
the corresponding trace below, the dephasing anisotropy is clearly
visible. Figure\ \ref{Fig3}a shows the SDTs, extracted from the
experimental data by fitting an exponentially damped cosine function
to the TRFR traces. The experimentally determined times are in
excellent agreement with the calculations. A rather striking feature
is the minimum in the SDT observed in Fig.\ \ref{Fig3}a for a
magnetic field of 0.2\ T, which is clearly present in both, the
experiment and the calculations. The corresponding experimental
trace is shown in Fig.\ \ref{Fig3}c, while Fig.\ \ref{Fig3}b shows
the trace at 0.2\ T for the perpendicular magnetic-field direction.
One can clearly see that in the latter situation the signal, and
hence the out-of-plane component of the spin polarization, reflects
more than half a precession cycle of the spin polarization during
the measured time window, while in the former case (Fig.\
\ref{Fig3}c), the spin polarization is completely relaxed within a
quarter cycle. Another feature is that there is a maximum of the SDT
at $B$=0.5\ T for a magnetic field along $[1\bar{1}0]$ direction.
The strength of the Rashba term, $\alpha=0.65\beta$,  in the
calculations is tuned to generate the best fit to the experimental
data, whereas the strength of the Dresselhaus term, $\beta=1.38$\
meV$\cdot$\AA, is determined following the work of Ref.\
\onlinecite{wu3}.\cite{comt1} The Rashba coefficient, $\alpha=0.9$ \
meV$\cdot$\AA,
 corresponds to a built-in electric field along the $z$-axis of
$16.8$\ kV/cm,\cite{Pfeffer} which is
  comparable with that of $15.0$\ kV/cm, which we calculated from the
  heterostructure of our sample self-consistently.\cite{Harrison}
Using these parameters, the SDTs for spins initially aligned
parallel to the magnetic field along the $[110]$ and $[1\bar{1}0]$
directions are calculated,\cite{comment} as Fig.\ \ref{Fig4}
shows. Here, we predict that the SDT for spins aligned
parallel to $[110]$ is several nanoseconds (black stars in
Fig.~\ref{Fig4}), comparable to the value observed in $[110]$-grown
quantum wells.\cite{Ohno}

The observed phenomena can be well understood by the inhomogeneous
broadening of the competing Dresselhaus and Rashba spin-orbit
couplings. As the inhomogeneous broadening is large/small in the
$[110]$/$[1\bar{1}0]$ direction,  the spin dephasing is large/small
for spin precession in the $(1\bar{1}0)$/$(110)$ plane, {\em i.e.},
along the $[1\bar{1}0]$/$[110]$ axis. As the spin-orbit field is
also strong,\cite{Stich} a small magnetic field can not force all
the spins to precess around it and there are spin precessions around
the direction perpendicular to the magnetic field. Therefore,
increasing the magnetic field suppresses the spin precessions along
the direction perpendicular to the magnetic field and hence
decreases/increases the SDT when the magnetic field is along the
$[110]$/$[1\bar{1}0]$ direction. On the other hand, if the magnetic
field is further increased, it will drive the spins to tilt toward
it in favor of the energy. Due to the anisotropy of the effective
magnetic field, the inhomogeneous broadening decreases/increases
when spins are tilted towards the $[110]$/$[1\bar{1}0]$ direction.
These two competing effects lead to the minimum/maximum of the SDT
observed in the experiment when the magnetic field is along
$[110]$/$[1\bar{1}0]$. We note that measurements with a very high
accuracy of alignment of the sample with respect to magnetic field
direction (less than 0.1 degree misalignment) indicate that for the
magnetic field values applied in this work, there is no significant
influence of a possible perpendicular field component due to slight
misalignment on the SDT.

In summary, we have performed time-resolved Faraday rotation
experiments on a high-mobility, $[001]$-grown 2DES. By applying a
magnetic field in two different inplane directions, we measure the
inplane anisotropy of the SDT. The measurements are compared to
the fully microscopic calculations and the comparison yields the
strength of the Rashba spin-orbit coupling. Using these
parameters, we calculate an inplane SDT of several nanoseconds for
our sample. We note that even though the TRFR experiment probes
the out-of-plane component of the spin polarization, it
nevertheless demonstrates the inplane spin dephasing anisotropy
and allows us to quantify it.

We gratefully acknowledge Jaroslav Fabian and R. T. Harley for
valuable discussions. This work was supported by the Deutsche
Forschungsgemeinschaft via GrK 638, grant No.\ Schu1171/1-3, SFB
689 and SPP1285. M.W.W. was supported by the Natural Science
Foundation of China under Grant No.\ 10574120, the National Basic
Research Program of China under Grant No.\ 2006CB922005, the
Knowledge Innovation Project of Chinese Academy of Sciences and
SRFDP. J.H.J. would like to thank J. Zhou for coding.

\end{document}